\documentclass[
reprint,
showpacs,
preprintnumbers,
nofootinbib,
bibnotes,
amsmath, amssymb,
aps,
prd,
floatfix,
]{revtex4-1}

\usepackage[pdftex]{graphicx}
\usepackage{dcolumn}
\usepackage{bm}
\usepackage{color}
\usepackage{multirow}
\usepackage{mathrsfs}
\usepackage{amssymb}
\usepackage{amsmath}

\begin{document}


\title{New Prospects in Fixed Target Searches for Dark Forces \\
with the SeaQuest Experiment at Fermilab}

\author{S.\ Gardner}
\email{\texttt{gardner@pa.uky.edu}}
\affiliation{Department of Physics and Astronomy, University of Kentucky,
Lexington, Kentucky 40506-0055 USA}

\author{R.\ J.\ Holt}
\email{\texttt{holt@anl.gov}}
\affiliation{Physics Division, Argonne National Laboratory, Argonne, IL 60439 USA}

\author{A.\ S.\ Tadepalli}
\email{\texttt{tadepalli@physics.rutgers.edu}}
\affiliation{Department of Physics and Astronomy, Rutgers University, Piscataway, NJ 08854 USA}




\begin{abstract}
An intense, 120 GeV proton beam incident on an extremely long,
iron target generates enormous numbers
of light-mass particles that also decay within that target.
If one of these particles decays to a final state
with a hidden gauge boson, or if such a particle is produced as a result of the initial
collision, then that weakly interacting,
hidden-sector particle may traverse the remainder of the target
and be detected downstream through its
possible decay to an $e^+e^-$, $\mu^+\mu^-$, or $\pi^+\pi^-$ final state.
These conditions can be realized through an extension of the SeaQuest
experiment at Fermilab, and in this initial investigation we
consider how it can serve as an ultrasensitive
probe of hidden vector gauge forces, both Abelian and non-Abelian.
A light, weakly coupled hidden sector may well explain the
dark matter established through astrophysical observations, and
the proposed search  can provide
tangible evidence for its existence --- or, alternatively,
constrain  a ``sea'' of possibilities.

\end{abstract}

\maketitle


\section{Introduction}
\label{sec:introduction}
Searches for new physics have long been motivated by the
 seeming inadequacies of the Standard Model (SM). Some are theoretical
and motivate searches for new physics at high-energy colliders, such as those that
would help
explain the origin of the weak scale,
$v=(2\sqrt{2}G_{F})^{-1/2}  \approx 174\,{\rm GeV}$~\cite{Veltman:1976rt,Susskind:1978ms,Baer:2006rs}.
Others include its inabilities to explain either
dark matter or dark energy and
their relative predominance over visible matter in the cosmic energy budget, as
deduced from astrometric observations~\cite{Bertone:2004pz,Feng:2010gw,Gardner:2013ama}.
The missing new physics can appear either at high energies, at short distance
scales~\cite{Jungman:1995df,Birkedal:2006fz,Gudnason:2006yj,Hooper:2007qk},
or at low energies, at
long distance scales~\cite{Okun:1982xi,Sikivie:1983ip,Moody:1984ba,Holdom:1986eq,Foot:2000vy,Dobrescu:2006au,Ahlers:2007rd,Ahlers:2007qf}.
Solutions to the dark matter
problem could conceivably come from
either source~\cite{Feng:2008ya,Kusenko:2009up,Feng:2010tg,Foot:2010av,Dall:2015bba}.
New, long-distance effects are both possible and discoverable
if the new, light degrees of freedom
couple to SM fields in a weak but yet appreciable way.
Such operators,
or portals, to possible hidden sectors have been discussed extensively in the
context of
apparent cosmic and gamma ray anomalies,
see Refs.~\cite{ArkaniHamed:2008qn,Cholis:2008hb,Fox:2008kb,Hooper:2010mq},
e.g., and have been proposed as an explanation
of the muon $g-2$ anomaly~\cite{Pospelov:2008zw}. (We also note earlier work in
which a light U(1) gauge boson directly couples to SM fermions to explain
the muon $g-2$~\cite{Fayet:1980rr,Gninenko:2001hx,Fayet:2007ua}, as well as
earlier astrophysical~\cite{Boehm:2003ha}, anomalies.)
In this paper we discuss the discovery prospects
of an ultrasensitive, broad-band search
for new, long-distance physics, made possible through
an extension of the SeaQuest E906 experiment at Fermilab.

The success of the SM in describing known particle phenomenology motivates
a framework in which new physics appears as additions to the SM
through effective operators ${\cal O}_i$ of mass dimension four
or higher.
The associated coupling constants
are characterized by $C_i/\Lambda^n$ where $n \ge 0$,
$C_i$ is dimensionless, and $\Lambda$ is the energy scale of new physics.
As we have noted,
these additions are thought to be either at high energy scales for which
$\Lambda > v$~\cite{Appelquist:1974tg,Buchmuller:1985jz,Grzadkowski:2010es}
with $C\sim {\cal O}(1)$ or at low energy scales for which $\Lambda \ll v$
or $n=0$ with $C \ll {\cal O}(1)$. In the latter category, the most effort has been
invested in operators for which $n= 0$, in part because
their appearance does not usually require the inclusion of additional
new physics to be theoretically consistent at high energy scales~\cite{Dall:2015bba}.
Dark photon searches fall into this class.
Higher-mass-dimension portals are also possible, but have received much
less attention.
Their coupling to SM particles is expected to be much smaller since $n>0$. The experiment
we consider, in which very small couplings can still be appreciable, thus serves
as an ideal hunting ground for such effects.
\begin{figure}
\includegraphics[clip,width=0.50\textwidth,angle=0]{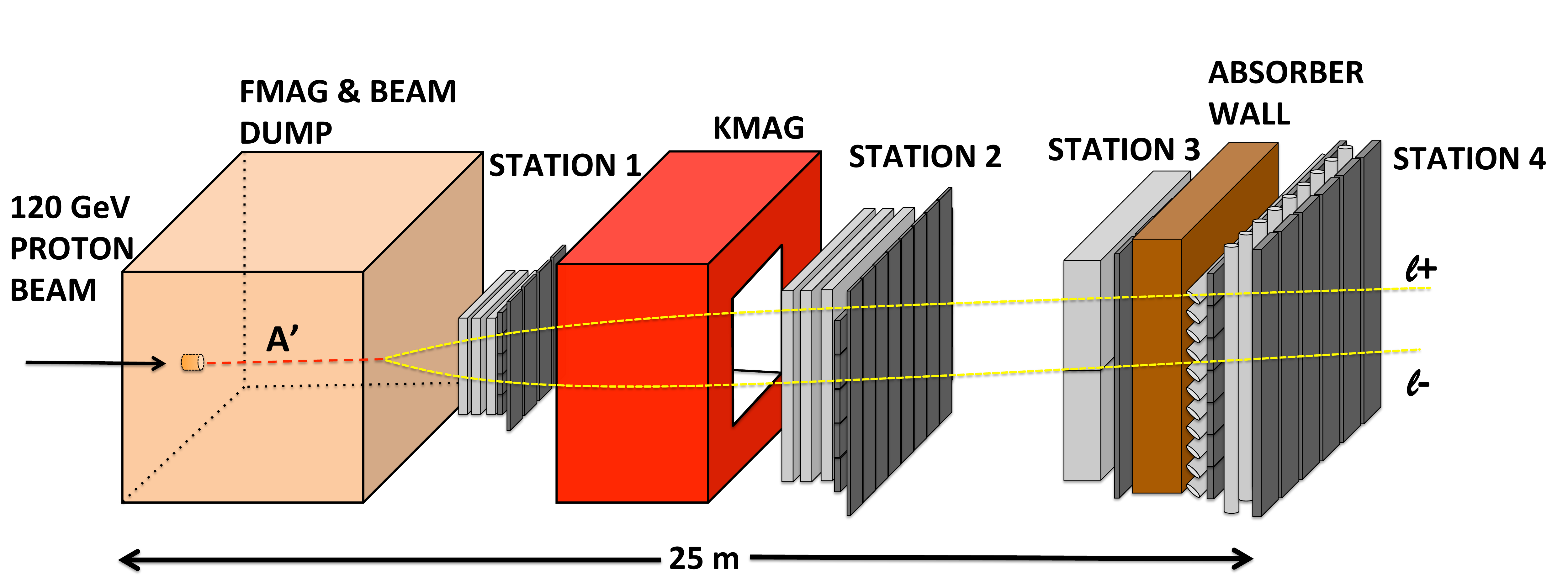}
\caption{\label{layout} Schematic of the SeaQuest spectrometer layout~\cite{Isenhower:2001zz}.
The 120 GeV proton beam from the Fermilab Main Injector
approaches the spectrometer through a 25-cm-long hole of 2.5 cm in diameter in the 5-m-long solid iron magnet.
An $A'$ generated in the first meter of the beam dump traverses the Focusing Magnet (FMAG) without being
affected by the magnetic field and can decay in the fiducial region into a lepton pair, or a pion pair (upon upgrade).
Stations 1, 2, and 3 comprise a series of drift chambers and an array
of scintillator hodoscope paddles used for track reconstruction and triggering purposes.
The 3-m-long air-gap KTeV Magnet (KMAG) is used to focus the muons back into
the spectrometer to facilitate momentum measurements.
The 1-m long iron absorber wall is followed by an array of proportional tubes used for muon identification.}
\end{figure}

Much experimental effort has been
invested in searches for rare exotic particles
through so-called beam-dump experiments, in which
detectors are mounted downstream of a
particle beam stopped in a target.
Our current investigation concerns the discovery prospects of
an experiment
of this class, and a schematic is shown in Fig.~\ref{layout}.
As emphasized by Ref.~\cite{Bjorken:2009mm},
the potential parameter space for a dark photon --- or, indeed, for any
particle probed via
a $n=0$ connector ---
is vast, in both candidate mass
and mixing angle.
Beam-dump experiments that involve
displaced vertices for dark photon  production and decay are largely
sensitive to small mixing angles, with electron and proton beam-dump
experiments giving comparable constraints~\cite{Bjorken:2009mm}.
Dedicated efforts have been
made to address the remaining holes in parameter
space~\cite{Bjorken:2009mm,Batell:2009di,Batell:2009yf},
with many recently completed searches and
reanalyses of earlier ones~\cite{Bjorken:1988as,Bjorken:2009mm,E141,Bross:1989mp,Gninenko:2011uv,Gninenko:2012eq,APEX_TEST,KLOE,Batley:2015lha,PHENIX,MAMI,Lees:2014xha,Blumlein:2011mv,Blumlein:2013cua,Gninenko:2013sr,Aad:2014yea,Aad:2015rba}
with many more
proposed and under
development~\cite{APEX_FULL,HPS,Freytsis:2009bh,Wojtsekhowski:2012zq,Beranek:2013yqa,Gninenko:2013rka,Andreas:2013lya,Echenard:2014lma,Alekhin:2015oba,LHCb}.

\begin{figure}
\includegraphics[scale=0.49]{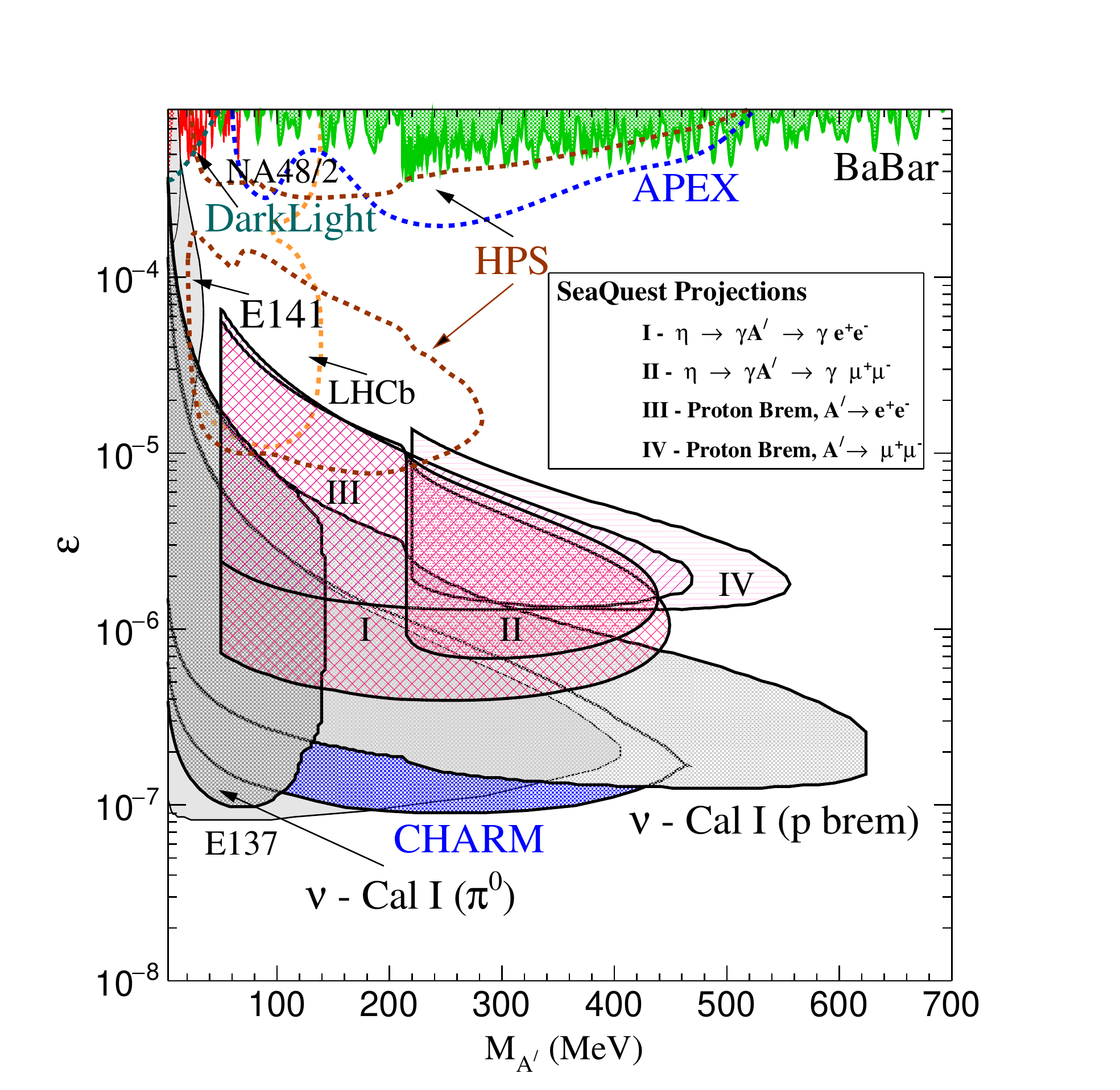}
\caption{\label{political}
Plot shows the projection contours of the coupling constant $\varepsilon$
as a function of dark photon mass $m_{A'}$ for four different processes that
could be used to search for dark photons at SeaQuest. Regions I and II are bounded
by the contour plots for $\eta \to \gamma A' \to \gamma e^{+}e^{-}$ and 
$\eta \to \gamma A' \to \gamma \mu^{+}\mu^{-}$, respectively, whereas regions III and IV 
refer to the limits inferred from use of the  proton bremsstrahlung production mechanism, 
followed by $A'\to e^+ e^-$ (III) and $A'\to \mu^+ \mu^-$ (IV) decay. 
The area excluded by electron beam dump experiments E137 \cite{Bjorken:1988as,Bjorken:2009mm},
E141 \cite{E141}, and the searches by BABAR \cite{Lees:2014xha}, CHARM \cite{Gninenko:2011uv,Gninenko:2012eq}, NA48/2 \cite{Batley:2015lha} are bounded by solid lines at 90 $\%$ CL, whereas those excluded by 
$\nu$-Cal I ($\pi^{0}$)\cite{Blumlein:2011mv} and $\nu$-Cal I (p-Brem)\cite{Blumlein:2013cua} are bounded by solid lines at 95 $\%$ CL. Also, the planned sensitivities of APEX (full run) \cite{APEX_FULL}, HPS~\cite{HPS}, DarkLight \cite{Freytsis:2009bh} (all at 90$\%$ CL) and LHCb\cite{LHCb} (at 95$\%$ CL) are shown as dotted lines for comparison. We omit the anticipated limits 
from VEPP-3~\cite{Wojtsekhowski:2012zq}, Refs.~\cite{Gninenko:2013rka,Andreas:2013lya}, 
Mu3e~\cite{Echenard:2014lma}, and MESA~\cite{Beranek:2013yqa}, which all probe 
lighter masses, as well as Ref.~\cite{Alekhin:2015oba}, for visual clarity. 
The region above $\varepsilon = 10^{-3}$ 
(not shown in the figure) has been excluded by several experiments such as E774 \cite{Bross:1989mp}, APEX (test run)\cite{APEX_TEST}, HADES \cite{Agakishiev:2013fwl}, KLOE \cite{KLOE}, PHENIX \cite{PHENIX}, MAMI \cite{MAMI}, 
along with the 2$\sigma$ exclusion limit obtained from $(g-2)_{e}$ \cite{Davoudiasl:2014kua}. 
Approximate limits at still weaker mixing angles from the LSND experiment \cite{Batell:2009di,Essig:2010gu,LSND_expt} and from astrophysical considerations \cite{Bjorken:2009mm,Dent:2012mx,Dreiner:2013mua,Fradette:2014sza,Foot:2014uba,Kazanas:2014mca} have been omitted. Note that the limits shown all assume that decays of the $A'$ to the invisible sector are nonexistent.}
\end{figure}

The extension of the SeaQuest experiment we consider
can contribute to this effort in different
ways.
Not only can it probe new regions of dark-photon parameter space,
leading either to a dark-photon discovery or a refinement of that phase space,
but it can also be used to probe dark forces that enter
solely through their mixing with QCD degrees of freedom.
In this latter case, proton beam-dump experiments play a special role, particularly
if the downstream spectrometer can detect pions.
In what follows we expound on the discovery prospects of the
experiment shown in Fig.~\ref{layout}. For reference, the
manner in which the existing SeaQuest spectrometer
can contribute to a dark photon
search, with reference to efforts under development worldwide,
is shown in Fig.~\ref{political}. We note that the SeaQuest experiment
can probe part of the dark-photon parameter space to be probed by the SHiP
experiment at CERN~\cite{Alekhin:2015oba}.
We also consider the discovery
prospects associated with a spectrometer upgrade to permit electron and pion detection.

We now sketch the content of the sections to follow. We begin with
an overview of hidden portal models, highlighting,
in particular,
the various ways in which quark and gluon degrees of
freedom can also play a role.
We then proceed to describe
the specific manner in which
hidden portals can be probed at SeaQuest. In this
initial investigation  we place a particular focus
on the radiative decays of the light mesons $\pi^0$ and $\eta$.
We expect light mesons to be produced copiously in a proton
beam dump experiment~\cite{Barabash:1992uq}, and their radiative decays are controlled
by the chiral anomaly even if the final state contains strongly
interacting particles.  The proton bremsstrahlung contour for SeaQuest in Fig.~\ref{political} was produced by following the method outlined in Ref.{~\cite{Blumlein:2013cua} and the simulation techniques described in sec. IV.A.
Finally, we turn to a discussion of the experimental
prospects, illustrating concretely how
SeaQuest can probe dark forces, before offering our
concluding summary.

\section{Hidden Sector Portals}
\label{sec:hidden}
The known interactions and particle content of the SM are richly complicated, and
this itself
suggests that the bulk of the matter content in the Universe should be similarly
complex. Existing constraints on its content, however, are minimal. Nevertheless,
the observation that dark matter is stable over at least Gyr time scales begs an
explanation, and it is natural to think that a gauge symmetry of the hidden sector can
provide it. The matter content of such a hidden sector need only interact
gravitationally with the matter we know --- no other interactions are required, though
it has become popular in recent years to build
theoretical models of cosmogenesis that tie the generation of dark
matter with that of the cosmic baryon asymmetry,
solving two problems at once~\cite{zurek,hoorabi,Servant:2013uwa}.
Although the existence of hidden sector gauge interactions
could potentially impact the
morphology of dwarf galaxies~\cite{Feng:2009mn} and have other
observational consequences~\cite{Kaplan:2009de,Cline:2012is,Fan:2013tia,Fan:2013yva},
the ability to discover such hidden sectors may rest on the
manner they can connect to the particles and interactions of the SM.

It has been popular to consider portals that consist of
operators that do not require new high-energy
physics for theoretical consistency, so that $n = 0$, or less.
An economical summary of such portals~\cite{Batell:2009di,Dall:2015bba}
is
\begin{equation}
{\cal L}_{\rm n\le 0}
= \kappa B^{\mu \nu} V_{\mu \nu} - H^\dagger H (A S + \lambda S^2) - Y_N L H N \,,
\label{master}
\end{equation}
where $B^{\mu \nu}$ is the field-strength
tensor associated with U(1)$_Y$ in the SM, $H$ is the SM Higgs field, and
$L$ is a SM left-handed (lepton) doublet.
The explicit hidden sector degrees of freedom are a field-strength tensor $V_{\mu \nu}$
associated with a hidden U(1) symmetry, a scalar $S$, and a fermion $N$. These new degrees of freedom
can couple to further hidden sector degrees of freedom,
and the latter can be richly diverse.

The new gauge boson degrees of freedom can be dark photons or $Z$'s,
or both, depending on the manner in which their mass is generated~\cite{Davoudiasl:2012ag,Davoudiasl:2014kua}.
If the dark photon is massless, kinetic mixing with the visible photon
would engender a DM electric charge~\cite{Holdom:1985ag}, and the resulting
constraints can be severe~\cite{Davidson:2000hf,Gradwohl:1992ue,Feng:2009mn}, though
they can also be weakened by either introducing multiple species with a net dark matter electric
charge of zero~\cite{Ackerman:2008gi} or by making the dark-matter particle a (nearly)
electrically neutral composite~\cite{Kaplan:2009de,Cline:2012is,CyrRacine:2012fz,Cline:2013pca}.
The latter, ``dark atom'' scenarios possess an additional length scale whose existence is
also constrained by cosmic microwave background (CMB) observations~\cite{Cyr-Racine:2013fsa}.
In this paper we focus on hidden vector gauge bosons that range from some 10 MeV to 700 MeV in mass, for
which the noted astrophysical constraints do not operate. We note in passing, however, that
extremely weakly coupled dark photons with such mass scales are nevertheless
constrained by supernova cooling~\cite{Bjorken:2009mm}, as well as
CMB and big-bang nucleosynthesis (BBN)~\cite{Fradette:2014sza}, considerations.
A minimal enlargement of the SM in the presence of
hidden gauge bosons is of form~\cite{Bjorken:2009mm,Batell:2009di,Batell:2009yf},
\begin{equation}
{\cal L} = {\cal L}_{\rm SM} +
\frac{1}{2}\kappa B^{\mu \nu} V_{\mu \nu}
- \frac{1}{4} V_{\mu \nu} V^{\mu \nu} + \frac{1}{2} m_{A'}^2 A'^{\mu} A'_\mu \,,
\end{equation}
where $V_{\mu \nu} =\partial_\mu A'_\nu - \partial_\nu A'_\mu$.
The kinetic mixing term $B^{\mu \nu} V_{\mu \nu}$ engenders, upon diagonalization
and field redefinition, a photon with a $A'$ admixture
controlled by the small parameter $\kappa$.
Specifically, $A^\mu \to A^\mu - \varepsilon A'^\mu$, where $\varepsilon \equiv \kappa \cos \theta_W$.
In constrast, mixing with the $Z$ is suppressed by a nominally large ratio of masses, namely
by ${\cal O}(\varepsilon m_{A'}^2/M_Z^2)$~\cite{Gopalakrishna:2008dv,Bjorken:2009mm,Davoudiasl:2012ag}. However,
with an enlarged Higgs sector,
such as in the two Higgs doublet model,
mixing with
both the photon and $Z$ can occur, and the candidate dark gauge boson becomes
a dark $Z$ or ``$Z_d$''~\cite{Davoudiasl:2012ag,Davoudiasl:2012ig,Davoudiasl:2012qa,Davoudiasl:2014kua}.
In this case, after diagonalization of the kinetic mixing term, the photon and $Z$ acquire a small admixture
of $Z_d$ which couples to SM electromagnetic and
weak-neutral currents as per~\cite{Davoudiasl:2012ag,Davoudiasl:2012ig,Davoudiasl:2012qa,Davoudiasl:2014kua}
\begin{equation}
{\cal L}_{\rm dark\, Z} =
- (\varepsilon e J_{\rm em}^\mu + \varepsilon_Z \frac{g}{2\cos \theta_W} J_{\rm NC}^\mu) Z_{d\,\mu} \,,
\end{equation}
with the earlier dark photon model emerging in the $\varepsilon_Z =0$ limit.
The appearance of $Z-Z_d$ mixing gives rise to low-energy
parity violating effects as well~\cite{Davoudiasl:2012ag,Davoudiasl:2012qa,Davoudiasl:2014kua}.
Nevertheless, it is apparent that dark photon searches also restrict the $Z_d$.

Thus far we have considered kinetic-mixing models associated with an Abelian gauge symmetry.
The gauge boson mass, be it that of $A'$ or $Z_d$, can be arranged through a Stueckelberg
mechanism~\cite{Stueckelberg:1938zz,Feldman:2007wj}, but
it can also be generated though a SM Higgs mechanism
in the hidden sector~\cite{ArkaniHamed:2008qn,Baumgart:2009tn}~\footnote{The hidden scalar can also function as a portal to the hidden sector, as per Eq.~(\ref{master}).}.
The discovery of a massive Abelian gauge boson
can thus implicitly hint to the existence of non-Abelian hidden-sector interactions as well.
Non-Abelian interactions can also appear explicitly, and
we now delve into this possibility directly and consider, in particular, how QCD degrees of freedom can
serve as a portal to a possible hidden sector.

It has long been thought that new matter with
 QCD-like interactions could exist~\cite{Okun:1980kw,Okun:1980mu,Gupta:1981ve}, though
the internal color symmetry of QCD would seem to preclude vector portals
of the sort we have considered thus far. Nevertheless, portals with QCD degrees of freedom need not
yield higher mass-dimension operators, nor do they necessarily require new UV degrees of freedom for theoretical
consistency. We highlight some of the possibilities in
what follows. We loosely organize our discussion in terms of portals of increasing mass dimension,
beginning with portals arising from a gauged baryon vector current and then turning to portals
employing multi-quark operators and gluons.

The possibility of a light U(1) gauge boson $B$ associated with a gauged baryon vector current
is a notion of some standing~\cite{Nelson:1989fx,Rajpoot:1989jb,Foot:1989ts,He:1989mi,Carone:1994aa,Carone:1995pu,Aranda:1998fr,FileviezPerez:2010gw,Tulin:2014tya,Dobrescu:2014fca},
though much of its focus has concerned the consequences
of its kinetic mixing
with the U(1)$_Y$ sector of the SM~\cite{Carone:1994aa,Carone:1995pu,Aranda:1998fr}.
Recently Tulin~\cite{Tulin:2014tya} has noted the possibility of a
mass-dimension-four connector in terms of quarks, namely,
\begin{equation}
{\cal L}_{\rm B} = \frac{1}{3} g_B {\bar q } \gamma^\mu q B_{\mu} \,,
\label{tulinL}
\end{equation}
where $B_\mu$ is the new gauge field that couples to baryon number. Since U(1)$_B$ is anomalous in the SM, its mixing with SM fields
requires new UV degrees of freedom for theoretical consistency~\cite{Nelson:1989fx,Tulin:2014tya,Dobrescu:2014fca,Dall:2015bba}, though such constraints have been
successfully implemented,
note, e.g., Refs.~\cite{Rajpoot:1989jb,Foot:1989ts,He:1989mi,FileviezPerez:2010gw,Dobrescu:2014fca}. Since the gauge
coupling $g_B$ is universal for all quarks,
the $B$ is taken to be strictly isoscalar, so that $B\to \pi^+\pi^-$ does not occur
if $G$ parity is not violated~\cite{Tulin:2014tya}. For $B$ bosons
in the mass window of $M_\pi \lesssim M_B \lesssim 620\,{\rm MeV}$ the primary decay
mode is thus $B\to \pi^0\gamma$.
Mixing of the $B$ with the photon
can appear through radiative corrections~\cite{Carone:1994aa,Carone:1995pu,Aranda:1998fr}, so that $B\to \ell^+\ell^-$
can occur as well. The $B$ can be probed through
the radiative decays of the light mesons~\cite{Nelson:1989fx,Tulin:2014tya}.
A distinct signature of this particular model is the appearance of
$B\to \pi^0\gamma$ decay, a final state which the SeaQuest spectrometer
cannot detect. Limits on the $A'$, however, translate to those on $B$ in a model-dependent way~\cite{Tulin:2014tya}.
It is worth noting, however, that the $B$ need not be strictly isoscalar.
As established from phenomenological studies of
the nucleon-nucleon force, both the $\rho$ and $\omega$ mesons couple to the baryon vector current
because the nucleon-nucleon force is not charge independent,
though the coupling of the $\omega$ is
roughly a factor of 5 larger
than that of the $\rho$~\cite{Machleidt:1987hj}.
Generalizing Eq.~(\ref{tulinL})
to the form
\begin{equation}
{\cal L}_{\rm B'} =  g^u_{B'} \bar u \gamma^\mu u B'_{\mu} + g^d_{B'} \bar d \gamma^\mu d B'_{\mu} + \dots \,,
\label{addL}
\end{equation}
we term this gauge boson $B'$ and relegate the contributions of other quarks to the ellipsis.
Thus $B'\to \pi^+\pi^-$ can occur without breaking $G$-parity, as had been assumed in
Ref.~\cite{Nelson:1989fx}. The $B'$ can be probed through the decay $\eta \to B' \gamma \to \pi^+\pi^- \gamma$,
which is accessible at SeaQuest. Other models
can be probed in this manner as well. For example, in the model of Dobrescu and Frugiuele~\cite{Dobrescu:2014fca},
the gauge boson that couples to baryon number is a leptophobic $Z'$,
where the interactions of the $Z'$ to quarks, for the first generation, is of form
\begin{equation}
{\cal L}_{\rm qZ'} = \frac{g_z}{2}
\left( z_{Q} {\bar Q}_L \gamma^\mu Q_L
+ z_{u} {\bar u}_R \gamma^\mu u_R + z_{d} {\bar d}_R \gamma^\mu d_R \right)Z_\mu'  \,.
\label{DFL}
\end{equation}
The field $Q_L$ is a left-handed quark doublet and $u_R$ and $d_R$ are right-handed singlets
--- we see that the $Z'$ can couple to isovector combinations of quarks as well.

We now turn to the possibility of portals comprised of operators of mass dimension greater than four.
The existence of such operators can
require the existence of new UV physics for theoretical consistency, but
this does not negate them as a possibility.
In this case it is possible to connect to non-Abelian hidden sector
degrees of freedom directly.
Starting in dimension six we note the following possibilities~\footnote{We note, however, that axion degrees of freedom can connect to SM fermions in dimension
five~\cite{Batell:2009di}, though their couplings to quarks are constrained by BBN~\cite{Blum:2014vsa}.}, namely,
\begin{equation}
\frac{\kappa^\prime}{\Lambda^2}
(\bar q_{L(R)} \gamma_\mu q_{L(R)}) (\bar q'_{L(R)} \gamma^\mu q'_{L(R)})
\end{equation}
where $q'$ is a hidden sector quark, admitting the possibility of a hidden sector ``pion''~\cite{Batell:2009di}, e.g.,
for which dark-matter models have recently been developed~\cite{Hochberg:2014dra,Hochberg:2014kqa}.
We note Refs.~\cite{Boddy:2014yra,Godbole:2015gma}
as alternate models of non-Abelian hidden sectors.
Explicitly non-Abelian portals also appear, such as~\cite{Baumgart:2009tn}
\begin{equation}
\frac{\kappa^\prime}{\Lambda^2}
{\rm tr}(\Phi^a F^a_{\mu\nu}){\rm tr}(\tilde\Phi^b {\tilde F}^{b\,\mu\nu})
\,,
\end{equation}
where $F^{a\,\mu \nu}$ is the QCD field strength tensor and
${\tilde F}^{b\,\mu\nu}$ denotes an analogous non-Abelian object,
though it need not reside in a theory of SU(3) color.
Note that the appearance of new heavy degrees of freedom is also explicit through that of the
heavy scalars $\Phi$ and $\tilde\Phi$
that transform under the adjoint representations of their respective groups~\cite{Baumgart:2009tn}.
Finally, in dimension eight it is also possible to have a pure glue-glue connector
\begin{equation}
{\rm tr}(F^{a\,{\mu\nu}} F^a_{\mu\nu})
{\rm tr}({\tilde F}^b_{\delta\rho} {\tilde F}^{b\,\delta\rho}) \,,
\end{equation}
which has also served as the basis for ``hidden valley'' models~\cite{Strassler:2006im} that can possess striking
collider signatures~\cite{Strassler:2006im,Han:2007ae,Kang:2008ea,Harnik:2011mv}.
This connector is also noteworthy as an UV ``complete'' extension of the SM. That is, operators
with mass dimension greater than four that break no SM symmetries do not require additional UV physics to be viable,
an idea exploited in the dark-matter model of Ref.~\cite{Godbole:2015gma}.

The various higher mass dimension connectors we have discussed appear in connection with QCD couplings, so
that ``infrared slavery''
at low energy can partially
offset the ratio of mass scale suppression associated with their higher mass dimension.
Indeed, at low energies we should recast the connectors we have considered in terms of
hadronic degrees of freedom.
A pertinent model is provided by
the hadronic kinetic mixing model of Ref.~\cite{Gardner:2013aiw}, which is based
on the hidden local symmetry model of
QCD~\cite{Bando:1984ej,Kitano:2011zk}, in which the $\rho$ mesons function
as effective gauge bosons of the strong interaction,
and on vector-meson dominance~\cite{Meissner:1987ge,O'Connell:1995wf}.
That is~\cite{Gardner:2013aiw},
\begin{eqnarray}
{\cal L}_{\rm mix} &=& -\frac{1}{4} \rho^a_{\mu\nu} \rho^{a\,\mu\nu}  -
\frac{1}{4} {\rho^{\prime\,a}_{\mu\nu}} \rho^{\prime\,a\,\mu\nu}  +
\frac{\varepsilon }{2} {\rho^a_{\mu\nu}} \rho^{\prime\,a\,\mu\nu} \nonumber \\
&+&
\frac{m_\rho^2}{2} \rho^a_\mu \rho^{a\,\mu} + \frac{m_{\rho^\prime}^2}{2}
\rho^{\prime\,a}_\mu \rho^{\prime\,a\,\mu}
+ \kappa_\rho J^{\mu\,a} \rho^a_\mu \,,
\label{mix}
\end{eqnarray}
where $J^{a\,\mu}$ denotes the hadronic vector current and
$\rho^{(\prime)\,a}$ are the gauge bosons of a hidden local SU(2) symmetry with
$\rho^{(\prime)\,a}_{\mu \nu} = \partial_\mu \rho^{(\prime)\,a}_\nu -
\partial_\nu \rho^{(\prime)\,a}_\mu$~\cite{O'Connell:1995wf}.
This model resembles the dark photon models discussed earlier but
contains two massive vector fields; its possible footprints at low energies, in beta decay,
have been considered by Gardner and He~\cite{Gardner:2013aiw}.
The kinetic mixing terms can be removed
through the field redefinition $\rho^a_\mu \to\rho^a_\mu - \varepsilon
\rho'^{a}_\mu$, thus yielding a coupling of
the hadronic vector current to $\rho'^a$.
Here the
$\rho'^0$ can be probed at SeaQuest.
Indeed all the quark-level models we have considered that
permit a hidden-sector contribution to a $\pi^+\pi^-$ final state can feed the
low-energy constant $\varepsilon$ in Eq.~(\ref{mix}). In this sense
a $\rho'$ search is rather generic, as it
constrains the $B'$, leptophobic $Z'$, and higher-dimension QCD
connectors we have considered as well.

\section{Probing Hidden Sectors at SeaQuest}
\label{sec:sea}

Enormous numbers of light mesons would be produced in the collision of an
intense, 120 GeV proton beam with a thick iron target, and we propose to search
for the light, hidden sector
particles that may be produced through their decays.
Moreover, dark gauge bosons
can be produced directly through initial-state radiation
from a high-energy proton beam, followed by a hard proton-iron collision; here we employ
the formalism of Ref.~\cite{Blumlein:2013cua}.

The models discussed in the previous section can all be probed
through an extension of SeaQuest, though the
distinctive features of a dark Z ---
its ability to mediate parity violation --- would not be apparent.
Nevertheless, sensitivity to a parity-violating dark gauge boson ($Z_d$)
should result from the use of a polarized proton beam~\cite{Isenhower:2012vh} via
the initial-state radiation production mechanism~\footnote{Parity violation could be
studied through use of either a polarized beam or an exceptionally long, polarized
target, but the latter is not planned for the extension of the SeaQuest
experiment we consider here.}. Specifically,
the $Z_d$ can be produced via either a vector or axial-vector coupling.
These processes can interfere to give different $\ell^+\ell^-$ yields as a function
of the beam helicity.

Irrespective of such developments,
searches for
hidden dark forces
in atomic parity violation, as well as in parity-violating
electron scattering,
play an important complementary
role~\cite{Davoudiasl:2012qa}.

The following decay chains are examples of
processes that can
be studied at SeaQuest. We note that $A'$ searches automatically limit the $Z_d$ as well.
\begin{itemize}
\item{\bf $\pi^0 \to \gamma A' \to \gamma  e^+ e^-$:} to search for $A'$.
\item{\bf $\eta \to \gamma A' \to \gamma \ell^+ \ell^-$ with $\ell \in e, \mu$:} to search for $A'$. The ability to detect different lepton final states offers the possibility of testing the
kinetic mixing mechanism.
\item{\bf $\eta \to \gamma A' \to \gamma \pi^+ \pi^-$:} to search for $A'$ --- or
$\rho'$, a ``strongly'' interacting hidden sector gauge boson.
The presence of the $\rho'$ would be signalled if no accompanying
events were observed in $\ell^+ \ell^-$.
\item{\bf proton bremstrahlung of $A'$ with $A'\to \ell^+,\ell^-$ and $\ell \in e, \mu$ or $A'\to \pi^+\pi^-$:} to search for $A'$.
Study of $A'\to \pi^+ \pi^-$ also permits a search for a leptophobic dark gauge boson $Z'$~\cite{Dobrescu:2014fca}.
\item{\bf Drell-Yan production of $A'$ with $A'\to \ell^+,\ell^-$ and $\ell \in e, \mu$:} to search for $A'$ --- studying
the possibility of $A'\to \pi^+\pi^-$ also opens sensitivity to a leptophobic dark gauge boson.
This perturbative QCD mechanism for dark gauge boson production is under study~\cite{pengprasad}.

\end{itemize}

Although we focus on the possibility of new vector gauge forces in this paper,
we wish to emphasize that the discovery prospects of the type of search we discuss
span a much broader horizon. We note that the study of light meson decays
in this context include these additional possibilities.

\begin{itemize}

\item{\bf $\eta \to \pi^0 h_d  \to \pi^0 \pi^+ \pi^-$:} to search for a
dark Higgs $h_d$, noting Refs.~\cite{Leutwyler:1990,Donoghue:1990xh},
as well as for scalar dark matter that mixes with QCD gluons~\cite{Godbole:2015gma}.

\item{\bf $K^\pm \to \pi^\pm Z_d \to \pi^\pm \ell^+ \ell^-$ with $\ell \in e, \mu$:} to search for $Z_d$~\cite{Davoudiasl:2012qa}.
\end{itemize}

We now
evaluate each of the radiative decay
processes in turn.

\subsection{The decay $\pi^0  \to \gamma A'\to \gamma \ell^+ \ell^-$}

The decay amplitude $\pi^0\to \gamma \gamma^{(*)}$ is controlled
by the primitive axial anomaly~\cite{Adler:1969gk,Bell:1969ts},
even if
QCD corrections that would alter its form only vanish
for on-mass-shell photons~\cite{Adler:1969er}.
Consequently, the partial width for Dalitz
decay is usually normalized relative
to that for $\pi^0 \to \gamma \gamma$~\cite{Dalitz:1951aj,Mikaelian:1972yg,Kampf:2005tz}.
To validate our procedures we begin
by computing
the partial width for Dalitz
decay at tree level
in the SM~\cite{Dalitz:1951aj,Mikaelian:1972yg,Kampf:2005tz},
which yields, after a partial integration over the three-body phase space,
\begin{eqnarray}
\Gamma(\pi^0 \to \gamma e^+ e^-) &=& \frac{2\alpha \Gamma_0}{3\pi}
\int_{r^2}^1 dx \,
{|f(1,0,x)|}^2
\frac{(1-x)^3}{x} \nonumber \\
&& \times \beta \left(1 + \frac{r^2}{2x}\right) \,,
\label{th}
\end{eqnarray}
where $\Gamma_0\equiv \Gamma(\pi^0 \to \gamma \gamma)$,
$x =s/M_{\pi^0}^2$, $r=2 m_e/M_{\pi^0}$, $\beta = \sqrt{1-r^2/x}$, and $s$ is the invariant mass of the
$e^+e^-$ pair. We note that $f(1,0,x)$ is the $\pi^0 \to \gamma \gamma^*$ transition
form factor (noting $f(p^2/M_{\pi^0}^2,k_1^2/M_{\pi^0}^2,k_2^2/M_{\pi^0}^2)$
for $\pi^0(p)\to \gamma^* (k_1) \gamma(k_2)$ decay~\cite{Mikaelian:1972yg}),
where
$f(1,0,0)=1$ appears in $\pi^0 \to \gamma \gamma$.
The transition form factor has been studied extensively
because of its connection to
the analysis of the hadronic-light-by-light contribution to the anomalous
magnetic moment of the muon, $g-2$,
and a model-independent representation, based on Pad{\'e} approximates,
of the transition form factor data from CELLO, CLEO, BaBar, and Belle, yields
\begin{equation}
f(1,0,x) = 1 + a_\pi x + b_\pi x^2 + {\cal O}(x^3)\,,
\label{piff}
\end{equation}
with $a_\pi = 0.0324 (12)_{\rm stat} (19)_{\rm sys}$
and $b_\pi = (1.06 (9)_{\rm stat} (25)_{\rm sys})\times 10^{-3}$~\cite{Masjuan:2012wy},
though a dispersive approach yields similar results with smaller errors:
$a_\pi = 0.0307 (6)$ and $b_\pi=(1.10 (2))\times 10^{-3}$~\cite{Hoferichter:2014vra}.
For reference an analysis of $\pi^0 \to \gamma e^+e^-$ in two-flavor
chiral perturbation theory (ChPT) enlarged
by electromagnetism yields $a_\pi =0.029 (5)$~\cite{Kampf:2005tz},
whereas ChPT at one loop with
$\mu = m_\rho$ yields $a_\pi =0.036$~\cite{Bijnens:1989jb}.
Computing
$\Gamma(\pi^0 \to \gamma e^+ e^-)/\Gamma_0$ from Eq.~(\ref{th})
using empirical inputs from Ref.~\cite{Agashe:2014kda}
yields $1.185\times 10^{-2}$
if the transition form factor is set to one,
in agreement with Refs.~\cite{Dalitz:1951aj,Kampf:2005tz}.
Including $f(1,0,x)$ using the
central values of Ref.~\cite{Masjuan:2012wy} yields $1.188\times 10^{-2}$, whereas
the experimental result $\Gamma(\pi^0 \to e^+ e^- \gamma)/\Gamma_0
=(1.188 (30))\times 10^{-2}$~\cite{Agashe:2014kda}.
We conclude that our
framework
works very well.
The three-body partial
width via an $A'$ intermediate state follows from inserting the factor
\begin{equation}
\frac{x^2}{\left(\left(x - x_{A'}\right)^2 + x_{A'}\tilde\Gamma_{A'}^2 \right)}
\, \varepsilon^4
\end{equation}
under the phase space integral, where $x_{A'} = {m_{A'}^2}/{M_{\pi}^2}$ and
$\tilde\Gamma_{A'} = {\Gamma_{A'}}/{M_{\pi}}$,
to yield
\begin{eqnarray}
&&\Gamma(\pi^0 \to \gamma A' \to \gamma e^+ e^- ) = \Gamma_0 \int_{r^2}^1 dx \, \frac{\alpha}{\pi} \varepsilon^4
| f(1,0,x)|^2 \nonumber \\
&&\times \frac{2(1-x)^3 x}
{\left(\left(x - x_{A'}\right)^2 + x_{A'}\tilde\Gamma_{A'}^2 \right)}
\beta \left(1 + \frac{r^2}{2x}\right) \,.
\label{thA}
\end{eqnarray}
Existing searches for hidden gauge bosons, such as $A'$ and $Z_d$,
have presumed that there are no lighter particles within the hidden sector, so that these particles
decay only to SM particles.
In this case this implies ${\cal B}(A' (Z_d) \to e^+ e^-)=1$.
Consequently, since $\tilde\Gamma_{A'} \sim \alpha \varepsilon^2 \ll 1$, we can employ the narrow width approximation,
replacing
\begin{equation}
\left(\left(x - x_{A'}\right)^2 + x_{A'}\tilde\Gamma_{A'}^2 \right)^{-1}
\longrightarrow \frac{\pi }{\sqrt{x_{A'}}\,\tilde\Gamma_{A'}} \delta\left(x - x_{A'} \right) \,.
\label{narrow}
\end{equation}
to yield
\begin{eqnarray}
\Gamma(\pi^0 \to \gamma A' \to \gamma e^+ e^- ) &=& | f(1,0,x_{A'})|^2
\Gamma(\pi^0 \to \gamma A')\nonumber \\
&&\times {\cal B}(A'\to e^+ e^- ) \,,
\label{pi2body}
\end{eqnarray}
where
\begin{equation}
\Gamma(\pi^0 \to \gamma A') = 2 \varepsilon^2 \left( 1 - \frac{m_{A'}^2}{M_{\pi}^2}
 \right)^3 \Gamma_0
\label{pi2gA}
\end{equation}
and
\begin{equation}
\Gamma(A' \to e^+ e^-) = \frac{1}{3} \alpha \varepsilon^2 m_{A'}
\sqrt{1 - 4 \frac{m_e^2}{M_{\pi}^2}}
\left( 1 + 2 \frac{m_e^2}{m_{A'}^2} \right) \,.
\label{pi2ee}
\end{equation}
We note $\Gamma_{A'} = \Gamma(A' \to e^+ e^-)$ by assumption.
As a result, Eq.~(\ref{pi2body}) can be viewed as the product of
two sequential two-body decays, in which the $A'$ appears on its mass shell,
moderated by the $\pi^0$ transition form factor. If, rather, $x_{A'}> 1$, then
the integral in Eq.~(\ref{thA}) is regular even if $\Gamma_{A'}=0$.
We set this possibility aside, however, both here and in what follows
because an on-mass-shell particle is needed to evaluate its transit through
the iron beam stop.
It is worth emphasizing that the transition form
factor we have noted in Eq.~(\ref{piff}) is developed
for the low-momentum transfer regime $x\ll 1$,
whereas $x_{A'}$ can be of ${\cal O}(1)$. Using a
Pad{\'e} approximate of $P_3^2(s)$ form~\cite{Masjuan:2012wy}\footnote{We thank P.~Masjuan for graciously providing it to us~\cite{Masjuan}.}, which
satisfies the asymptotic
limit from perturbative QCD~\cite{Lepage:1979zb} for the transition form factor,
we have checked that for our application
its numerical effects are
so slight
that we can neglect it with impunity.

\subsection{The decay $\eta  \to \gamma A'\to \gamma \ell^+ \ell^-$}

To evaluate $\eta\to \gamma \ell^+ \ell^-$ decay in the SM we need
only replace $M_{\pi} \to M_{\eta}$ and $m_e \to m_\ell$
in Eq.~(\ref{th}) and modify the transition form factor. A
new parametrization~\cite{Escribano:2015nra} of the
$\eta \to \gamma^\ast \gamma$ form factor, including the latest measurement of
$\eta\to e^+ e^- \gamma$~\cite{Aguar-Bartolome:2013vpw}, and
employing rational approximates~\cite{Escribano:2013kba},
has recently become available.
With $x \to s/M_{\eta}^2$, and as $x\to 0$ (with $x>0$) we have
\begin{equation}
f_\eta (1,0,x) = 1 + b_\eta x + c_\eta x^2 + d_\eta x^3 + {\cal O}(x^4)\,,
\label{etaff}
\end{equation}
where $b_\eta = 0.576 (11)_{\rm stat}(4)_{\rm sys}$, $c_\eta = 0.339 (15)_{\rm stat}(5)_{\rm sys}$,
and $d_\eta = 0.200 (14)_{\rm stat}(18)_{\rm sys}$~\cite{Escribano:2015nra}.
Since the integral of Eq.~(\ref{th}) includes $x=1$ as well, we
also evaluate $f_\eta (1,0,x)$ using the complete rational approximate\footnote{We
thank the authors of Ref.~\cite{Escribano:2015nra}
for a high precision version of the function
$P_1^7(Q^2)$ that appears in their Table 5,
making it suitable for numerical work~\cite{Escribano:2015nra,Masjuan}.}, noting that
time-like data is available up to $s\approx 0.22 \,{\rm GeV^2}$~\cite{Escribano:2015nra}.
Employing Eq.~(\ref{th}) and Eq.~(\ref{etaff}) to compute
$\Gamma(\eta \to \gamma e^+ e^-)/\Gamma_0^\eta$ yields
 $1.67\times 10^{-2}$ (and $1.62\times 10^{-2}$ without the form factor)
 to compare with the experimental ratio $1.75 (06) \times 10^{-2}$~\cite{Agashe:2014kda}.
Computing, rather, $\Gamma(\eta \to \gamma \mu^+ \mu^-)/\Gamma_0^\eta$ yields
 $8.17\times 10^{-4}$ (and $5.51\times 10^{-4}$ without the form factor)
 to compare with the experimental ratio $7.87 (13) \times 10^{-4}$~\cite{Agashe:2014kda}.
Using the complete rational approximate makes for little impact, changing
only the value of $\Gamma(\eta \to \gamma \mu^+ \mu^-)/\Gamma_0^\eta$
result to $8.19 \times 10^{-4}$
once the empirical central value of
$\Gamma_0^\eta\equiv \Gamma(\eta \to \gamma\gamma)$ is imposed.
The estimates we have made agree with the experimental central values
at the $\approx 5\%$ level, a disagreement
larger than one would naively expect
from radiative corrections but which, rather, appears to stem from differences
in recent experimental results for
$\Gamma(\eta\to e^+e^- \gamma)$~\cite{Berghauser:2011zz,Aguar-Bartolome:2013vpw}
and the world average~\cite{Agashe:2014kda}. It is worth noting that the
$\eta$ decay partial widths are measured relative to $\Gamma_0^\eta$, and
although a recent, precision measurement of $\Gamma_0^\eta$ exists~\cite{Babusci:2012ik},
differences in measurements of $\Gamma_0^\eta$ from the $e^+e^-$ collisions and
electron-nucleus scattering, through use of the Primakoff effect, impact earlier
determinations of the partial widths~\cite{Agashe:2014kda}.
Irrespective of this, we conclude that
our framework certainly works sufficiently well for our current
purpose. Turning to the possibility of an $A'$ intermediate state, we
update Eq.~(\ref{thA}) to
\begin{eqnarray}
&&\Gamma(\eta \to \gamma A' \to \gamma \ell^+ \ell^- ) = \Gamma^\eta_0
\int_{r_\ell^2}^1 dx \, \frac{\alpha}{\pi} \varepsilon^4
| f_\eta(1,0,x)|^2 \nonumber \\
&&\times \frac{2(1-x)^3 x}
{\left(\left(x - x_{A'}\right)^2 + x_{A'}\tilde\Gamma_{A'}^2 \right)}
\beta \left(1 + \frac{r_\ell^2}{2x}\right) \,,
\label{etathA}
\end{eqnarray}
where $M_{\pi^0} \to M_{\eta}$, $r_\ell=2 m_\ell/M_\eta$, and
$\beta = \sqrt{1-r_\ell^2/x}$. Here, too, we can use the narrow
width approximation to find
\begin{eqnarray}
\Gamma(\eta \to \gamma A' \to \gamma \ell^+ \ell^- ) &=& | f_\eta(1,0,x_{A'})|^2
\Gamma(\eta \to \gamma A')\nonumber \\
&&\times {\cal B}(A'\to \ell^+ \ell^- ) \,,
\label{eta2body}
\end{eqnarray}
where $\Gamma(\eta \to \gamma A')$ and $\Gamma(A'\to \gamma \ell^+ \ell^- )$
follow from suitable substitutions in Eqs.~(\ref{pi2gA}) and (\ref{pi2ee}).

\subsection{The decay $\eta  \to \gamma A'\to \gamma \pi^+ \pi^-$}

To study this particular channel we need only compute $\Gamma(A' \to \pi^+\pi^-)$, replacing
$\ell$ with $\pi$ in Eq.~(\ref{eta2body}). Recalling the definition of the $\pi$ form factor $F_\pi (s)$
in $e^+e^-\to \pi^+ \pi^-$ at center-of-mass energy $\sqrt{s}$~\cite{Gardner:1997ie}, we note the decay amplitude is of form
\begin{equation}
\!\!\!\!\!{\cal M}(A'(p) \to \pi^+(p_1)\pi^-(p_2)) =  \varepsilon e \varepsilon_{A'}^\mu (p_1 -p_2)_\mu F_\pi(p^2)\,,
\end{equation}
so that the decay width is
\begin{equation}
\Gamma(A'\to \pi^+\pi^-) = \frac{1}{12} \varepsilon^2 \alpha m_{A'} \left(1 - 4 \frac{M_\pi^2}{m_{A'}^2} \right)^{3/2} \!\!|F_\pi(m_{A'}^2)|^2 \,.
\label{Ap2pi}
\end{equation}
This can be compared to the estimate used earlier in the literature~\cite{Bjorken:2009mm,Blumlein:2013cua}
\begin{equation}
\Gamma(A'\to {\rm hadrons}) = \frac{1}{3} \varepsilon^2 \alpha m_{A'} {\cal R}(m_{A'}^2)
\label{Ap2had}
\end{equation}
where the cross-section ratio
\begin{equation}
{\cal R}(s) \equiv
\frac{\sigma(e^+e^- (s) \to {\rm hadrons})}{\sigma(e^+e^- (s) \to \mu^+ \mu^-)}
\end{equation}
below $\pi \omega$ threshold
evaluates to
\begin{equation}
{\cal R}(s) = \frac{1}{4} \frac{(s - 4 M_\pi^2)^{3/2} |F_\pi(s)|^2}{(s - 4 M_\mu^2)^{1/2} (s+ 2 M_\mu^2)}\,,
\label{map}
\end{equation}
revealing that Eq.~({\ref{Ap2had}) is a very good
approximation over a large mass range
--- only the factor $(1- 4 M_\mu^2/s)^{1/2} (1+2 M_\mu^2/s)$ is extraneous.
A simple form of $F_\pi$ especially
suitable for $\sqrt{s} \lesssim 600\,{\rm MeV}$ is~\cite{Harrison:1998yr}
\begin{equation}
F_\pi(s) = \frac{-f_{\rho\gamma}g_\rho}{s - M_\rho^2 + i \Pi(s)}\,,
\label{fpi}
\end{equation}
where $\Pi(s) = (M_\rho^2/\sqrt{s})(p(s)/p(M_\rho^2))^3 \Gamma_\rho$ includes a
running $\rho$ width, with $p(s)=\sqrt{s/4 - M_\pi^2}$ and
a normalization as per Ref.~\cite{Gardner:2001gc}.
An explicit test of this form is shown in Fig.~5 of
Ref.~\cite{Gardner:2001gc},
for which $f_{\rho\gamma}=0.122(1)\,{\rm GeV}^2$~\cite{Gardner:1998ta},
$g_\rho=5.8$, $M_\rho=769.3\,{\rm MeV}$ and $\Gamma_\rho=150\,{\rm MeV}$
are used. The $\rho$ is a broad resonance, and if its mass and width are determined as per
the real-axis prescription of Eq.~(\ref{fpi}),
with a $s$-dependent width, as distinct from the location of its pole
in the complex plane, then the resulting resonance parameters depend on the
precise form of $F_\pi(s)$~\cite{Gardner:1997ie,Gardner:1998ta}.
Such a real-axis prescription is employed in the
determinations averaged by Ref.~\cite{Agashe:2014kda}, for which
$M_\rho=775.26(25)\,{\rm MeV}$ and $\Gamma_\rho=149.1(8)\,{\rm MeV}$ are reported.
We employ this latter set of parameters in our analysis here, and the incurred differences
are negligible. We display the resulting $A'$ branching ratios in Fig.~\ref{Apbranch}, and
the results are similar, though
our $\Gamma(A'\to \pi^+\pi^-)$ is a bit larger,
than those determined using the
prescription of Eq.~(\ref{Ap2had})~\cite{Blumlein:2013cua}.
\begin{figure}
\includegraphics[clip,width=0.50\textwidth,angle=0]{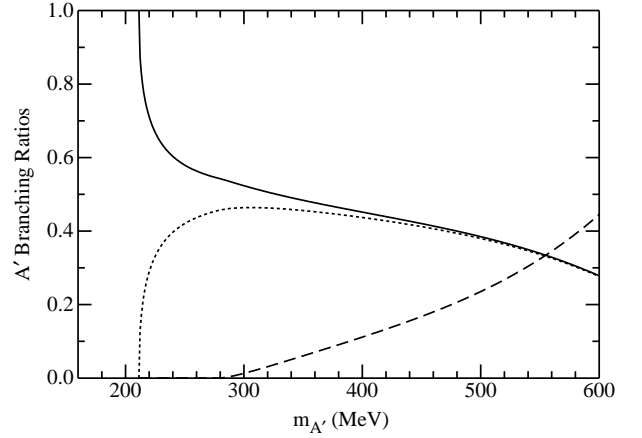}
\caption{\label{Apbranch}
The branching ratios for ${\cal B}(A'\to e^+ e^-)$ (solid), ${\cal B}(A'\to \mu^+ \mu^-)$ (dotted), 
and ${\cal B}(A'\to \pi^+ \pi^-)$ (dashed)
as per Eqs.~(\ref{pi2ee}), (\ref{Ap2pi}), and (\ref{fpi}), assuming that
decays to the hidden sector do not occur,
as a function of the $A'$ mass in GeV.
}
\end{figure}

\subsubsection{On $\Gamma(A'\to {\rm hadrons})$ at higher energies}
The proton initial-state radiation mechanism for dark
gauge boson production can probe $A'$ masses much in excess
of 600 MeV, so that we now develop an expression for
$\Gamma(A'\to \pi^+ \pi^-)$, as well as for
$\Gamma(A'\to {\rm hadrons})$,
in this mass region. To go beyond
the prescription of Eq.~(\ref{Ap2had}) in
estimating $\Gamma(A'\to {\rm hadrons})$ we use
Eq.~(\ref{map}) at $s=m_{A'}^2$ to rewrite Eq.~(\ref{Ap2pi}) as
\begin{eqnarray}
\Gamma(A'\to \pi^+\pi^-) &=& \frac{1}{3} \varepsilon^2 \alpha m_{A'}
\left(1 - 4 \frac{M_\mu^2}{m_{A'}^2} \right)^{1/2} \nonumber \\
&\times& \left(1 + 2 \frac{M_\mu^2}{m_{A'}^2} \right)
{\cal R}(m_{A'}^2) \,,
\label{BRnew}
\end{eqnarray}
which is some 10\% smaller than Eq.~(\ref{Ap2had}) if $m_{A'}=300\, {\rm MeV}$.
For $m_{A'}$ in excess of $\pi\omega$ threshold, or $\approx 920\, {\rm MeV}$,
Eq.~(\ref{BRnew}) is simply that of $\Gamma(A'\to {\rm hadrons})$,
implying that ${\cal B}(A'\to e^+ e^-) + {\cal B}(A'\to \mu^+ \mu^-) +
{\cal B}(A'\to \pi^+ \pi^-) < 1$ as well.

\subsubsection{On $\Gamma(A'\to {\rm invisibles})$}

It is entirely possible that a produced dark gauge boson also
decays invisibly to the particles of the hidden sector, though
this possibility has been neglected in early studies~\cite{Bjorken:2009mm}.
Nevertheless, it
can readily be incorporated, both in this case and throughout,
by making the total visible decay width some fixed fraction of the total width.
Varying the fraction of decays to the invisible sector consequently changes
the exclusion plot.

\subsection{The decay $\eta \to \gamma \rho' \to \gamma \pi^+ \pi^-$}

Now we turn to the possibility of a dark gauge boson, the $\rho'$,
which may or may not be a composite object.
To anchor our analysis we first compute the rate for
$\eta (p) \to \gamma (k)  \pi^+ (p_2) \pi^- (p_1)$ in the SM, where
we assume, in analogy to
our earlier work, that
 the decay vertex is mediated by the
axial anomaly in the low-energy,
chiral theory of hadrons~\cite{Wess:1971yu,Witten:1983tw}, where
we refer to Ref.~\cite{Bijnens:1989jb} for an analysis of
this assertion within chiral perturbation theory.
In particular, we assume that the decay is dominated by $\eta\to (\pi\pi)_{L=1} \gamma$
channel with the $\rho$ playing a prominent role.
Since $\rho \to \pi^+ \pi^-$ is a $p$-wave decay, we define the coupling $g_\rho$ as per
\begin{equation}
\langle \pi^+ (p_2) \pi^- (p_1) | \rho^0 (p_\rho, \varepsilon_\rho)
\rangle = - g_\rho \varepsilon_\rho \cdot (p_2 - p_1)
\label{rhocoup}
\end{equation}
to yield the decay amplitude
\begin{equation}
{\cal M} = - \frac{A_\eta}{2} \varepsilon_{\mu \nu \rho\sigma} \varepsilon^{\mu \, *}(k) \varepsilon_{\rho}^{\nu \, *}(p_{\rho})
k^{\rho} p_{\rho}^\sigma \frac{f_{\rho\gamma} g_{\rho} \varepsilon_\rho \cdot (p_2 - p_1)}{s - M_\rho^2 + i \Pi(s)} \,,
\end{equation}
where $s=p_\rho^2$ and $\Pi(s)$ is defined after Eq.~(\ref{fpi}).
Summing over the $\rho$ spin,
this becomes
\begin{equation}
{\cal M} =  - A_\eta \varepsilon_{\mu \nu \rho\sigma} \varepsilon^{\mu \, *}(k) p_1^\nu
k^{\rho} p_2^\sigma \frac{f_{\rho\gamma} g_{\rho}}{s - M_\rho^2 + i \Pi(s)} \,.
\end{equation}
We note that this is just a special case of
${\cal M}= \varepsilon_{\mu \nu \rho \sigma} \varepsilon^{\mu \, *} p_1^\nu k^\rho p_2^\sigma P(s) F_\pi (s)$,
where
$P(s)$ is a polynomial of form $P(s)=A_\eta(1 + \alpha s)$~\cite{Stollenwerk:2011zz}.
The corrections to this framework in $\eta \to \pi^+\pi^- \gamma$ decay
are very small~\cite{Stollenwerk:2011zz,Hanhart:2013vba,Kubis:2015sga},
supporting its use in the analysis here.
Evaluating
the integral over the three-body
phase space (with the integral over solid angle computed in the rest frame of the
$\pi^+ \pi^-$ pair) yields
\begin{eqnarray}
\Gamma(\eta\to\gamma\pi^+\pi^-) &=& \frac{M_\eta^7 |A_\eta|^2}{3\pi^3 2^{11}}
\int_{r^2}^1 dx\, x (1-x)^3 (\sqrt{ 1 - r^2/x})^3 \nonumber\\
&&\times |F_\pi(x M_{\eta}^2)|^2 (P_\eta(x M_\eta^2))^2\,,
\label{decay}
\end{eqnarray}
where $r=2 M_\pi/M_\eta$, $x= s/M_\eta^2$, and we have included the polynomial
of Ref.~\cite{Stollenwerk:2011zz} as $P_\eta(s) = 1 + \alpha s$ with
$\alpha= (1.96 (27)_{\rm stat} (2)_{\rm sys}) \, {\rm GeV}^{-2}$~\cite{Stollenwerk:2011zz}.
Although we could determine the constant $A_\eta$ by simply appealing to
the empirical $\eta \to \gamma \pi^+ \pi^-$ partial width, it is
worth noting that the
axial anomaly can fix this constant as well.
That is, the low-energy theorem relating $\gamma^*\to\pi^+\pi^-\pi^0$
to $\pi^0\to \gamma\gamma$~\cite{Adler:1971nq,Aviv:1971hq,Terent'ev:1971kt}
generalizes in the chiral SU(3)$_f$ limit at zero momenta to yield
$A_\eta = e /(4\sqrt{3} \pi^2 f_\pi^3)$~\cite{Venugopal:1998fq,Holstein:2001bt}.
Using $f_\pi=92.2\,{\rm MeV}$, this yields $A_\eta=5.65\,{\rm GeV}^{-3}$,
whereas including SU(3)$_f$ breaking and $\eta-\eta'$ mixing gives
an additional multiplicative factor of
$\xi=\cos \theta f_\pi/f_8 - \sin \theta \sqrt{2} f_\pi/f_0$~\cite{Venugopal:1998fq,Holstein:2001bt},
which evaluates to $\xi=1.057$ using $f_8/f_\pi=1.3$, $f_0/f_\pi=1.04$, and
$\theta=-20^\circ$~\cite{Venugopal:1998fq,Holstein:2001bt}.
Doing the integral in Eq.~(\ref{decay}) using Eq.~(\ref{fpi}),
$A_\eta=5.65\,{\rm GeV}^{-3}$, and
$\Gamma_{\rm tot}=1.31 (5)\,{\rm keV}$~\cite{Agashe:2014kda}
yields a branching ratio of $6.42\%$, whereas
the experimental result is
${\cal B}(\eta\to \pi^+\pi^- \gamma)= (4.22 (8))\%$~\cite{Agashe:2014kda}, a fractional
difference of some $30\%$.
Including $\xi$ and an overall
correction factor to $A_\eta$ of $(1+\delta)$ with $\delta=-0.22 (4)$
from the assessment of Ref.~\cite{Stollenwerk:2011zz}, yields
${\cal B}(\eta\to \pi^+ \pi^- \gamma)=4.37\%$, in reasonable agreement with
experiment given the various uncertainties. To adapt this framework to the computation of
$\Gamma(\eta \to \gamma \rho' \to \gamma \pi^+ \pi^-)$ we need only replace
$(s - M_\rho^2)^2 + \Pi^2(s)$ in Eq.~(\ref{fpi}) by
$(s - M_{\rho'}^2)^2 + M_{\rho'}^2 \Gamma_{\rho'}^2$
and multiply by $\varepsilon^4$. Although ${\cal B}(\rho'\to\pi^+\pi^-)$ could
plausibly be less than one, we shall assume that it is still reasonable to
use the narrow width approximation of Eq.~(\ref{narrow}), yielding
\begin{eqnarray}
&&\Gamma(\eta\to \gamma \rho' \to \gamma \pi^+\pi^-) =
\frac{\varepsilon^4}{3 \pi^2 2^{11}} M_\eta^3 |A_\eta (1+\delta)|^2
 \frac{M_{\rho'}}{\Gamma_{\rho'}} \nonumber \\
&&\times P_{\eta}^2(M_{\rho'}^2) f_{\rho \gamma}^2 g_{\rho}^2
\left(1 - \frac{M_{\rho'}^2}{M_{\eta}^2}\right)^3 \left(1 - \frac{4 M_{\pi}^2}{M_{\rho'}^2}\right)^{3/2} \,.
\end{eqnarray}
Supposing a $\rho'$ coupling of the form given in Eq.~(\ref{rhocoup}),
but with strength $g_\rho \varepsilon$, we have
\begin{equation}
\Gamma(\rho' \to \pi^+\pi^-) = \frac{1}{48\pi} \varepsilon^2 g_\rho^2 M_{\rho'} \left( 1 - \frac{4M_\pi^2}{M_{\rho'}^2} \right)^{3/2} \,.
\label{rhop2pi}
\end{equation}
Thus if $\Gamma_{\rho'} = \Gamma(\rho' \to \pi^+ \pi^-)$, we find
\begin{eqnarray}
\Gamma(\eta\to \gamma \rho') &=& \frac{\varepsilon^2}{128 \pi} M_\eta^3 |A_\eta (1+\delta)|^2 P_{\eta}^2(M_{\rho'}^2) f_{\rho \gamma}^2
\nonumber \\
&&\times {\left(1 - \frac{M_{\rho'}^2}{M_{\eta}^2}\right)^3 }\,.
\label{decay2rhop}
\end{eqnarray}
In what follows we include the possibility of invisible decays by studying
the impact of making ${\cal B}(\rho'\to\pi^+\pi^-)$ some definite fraction
less than one on the parameter exclusion plot.

\section{Experimental Prospects}

Already, there are a large number of searches for dark photons and a number of
planned searches~\cite{Bjorken:1988as,Bjorken:2009mm,E141,Bross:1989mp,Gninenko:2011uv,Gninenko:2012eq,APEX_TEST,KLOE,Batley:2015lha,PHENIX,MAMI,Lees:2014xha,Blumlein:2011mv,Blumlein:2013cua,Aad:2014yea,Aad:2015rba,APEX_FULL,HPS,Freytsis:2009bh,Wojtsekhowski:2012zq,Beranek:2013yqa,Gninenko:2013rka,Andreas:2013lya,Echenard:2014lma,Alekhin:2015oba,LHCb}.

We illustrate here a possible new search for dark photons at the FNAL E906 (SeaQuest) experiment at Fermilab.  This experiment was designed to perform a measurement of Drell-Yan processes for a proton beam incident on stationary targets.  The experiment is comprised of an intense ($5\times10^{12}$
protons/spill at 1 spill per min) 120 GeV proton beam incident on a target.
Less than a meter downstream of the target is a 5-m long solid iron magnet that not only serves to begin the analysis of the muon pair from the
Drell-Yan process, but also as a beam dump.
This beam dump tremendously attenuates all hadrons and most charged particles from the target with the exception of muons.
Following the solid iron magnet, there are two stations of scintillator hodoscopes and drift chambers that can define the vertex of either a prompt Drell-Yan
event in the target or beam dump or from the decay of a prompt or displaced vertex from a dark photon.
The first detector station is followed by an air gap magnet and two additional detector stations.
This system serves as the pair spectrometer and can measure the mass of an event with a relative mass resolution that
varies between 2.5-6\% depending on where the event occurs in the solid iron magnet.
A 1-m thick block of iron was placed just upstream of the fourth detector station and serves to filter most charged particles except muons.
The fourth detector substation comprises a layer of proportional tubes and scintillator paddles that serve as a muon identification system.
A schematic overview of the SeaQuest spectrometer is shown in Fig.~\ref{layout}~\cite{Isenhower:2001zz}.

Clearly, experiment E906 has all the basic elements of a ``shining-through-the-wall" beam stop experiment to search for dark photons.  The basic elements consist of a high-energy proton beam incident on a fixed target and a pair spectrometer which detects lepton pairs from the decay of the dark photon, where the spectrometer is well-shielded from the target in order to minimize background events from ordinary SM processes.  Consequently, the dark photons, being weakly interacting, can travel unscathed through the shield where they decay into lepton pairs and are subsequently detected in the pair spectrometer.

We have been conservative in the following simulations.  We have only considered events where the background is known to be low, namely after all or most of the shield.  For the special case of muon pair detection, we have used part of the last meter of the Fe shield as part of the fiducial region since the Fe absorber has little effect on the muons.  As more is known about the experiment and if more trigger optimization can be implemented, it may be possible to extend the fiducial region further upstream in the beam dump.
However, such studies are beyond the scope of this work.

\subsection{$\eta$ meson decay to charged lepton pairs}

Here we illustrate the regions of sensitivity in parameter space of $\varepsilon$ and dark photon mass, $m_{A'}$,
for SeaQuest for four different processes that could be used to search for dark photons. Fig.~\ref{political}
shows the projected regions of sensitivity for SeaQuest along with
areas excluded by electron and proton beam dump experiments and proposed experiments.
The simulation assumes that we have a 200-day experiment with an overall efficiency of about 30\%
which approximates the conditions experienced in the experiment thus far. We investigated dark photon production from radiative $\eta$ and $\pi^{0}$ decays and proton bremsstrahlung. The $\eta$ and $\pi^{0}$ yields per proton have been estimated from GEANT4 simulations that record the energy and transverse momentum spectra of  $\eta$ and $\pi^{0}$ mesons produced (shown in Fig.~\ref{eta_energy} and Fig.~\ref{eta_momentum}) when 6.2 million 120 GeV protons are incident on a 4.75-m long Fe beam dump. These energy and $p_{T}$ distributions were used in the calculation of the estimation of the dark photon yield.

\begin{figure}[ht]
\includegraphics[clip,width=0.50\textwidth,angle=0]{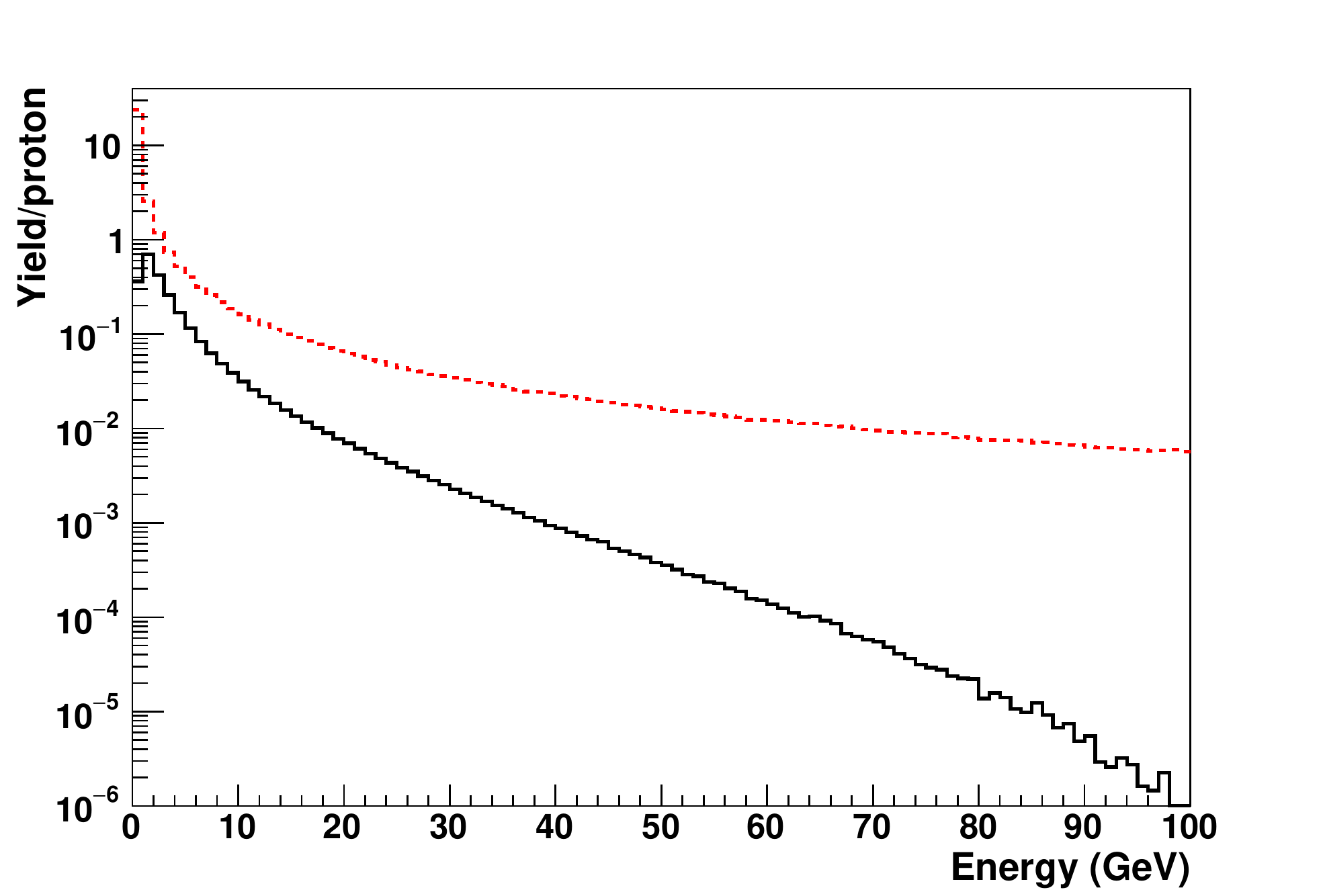}
\caption{\label{eta_energy} The $\eta$ (solid) and $\pi^{0}$ (dotted) yield/proton as a function of the energy of the particles as obtained from GEANT4 Monte Carlo simulations of 120 GeV protons interacting with the Fe beam dump.}
\end{figure}

\begin{figure}[ht]
\includegraphics[clip,width=0.50\textwidth,angle=0]{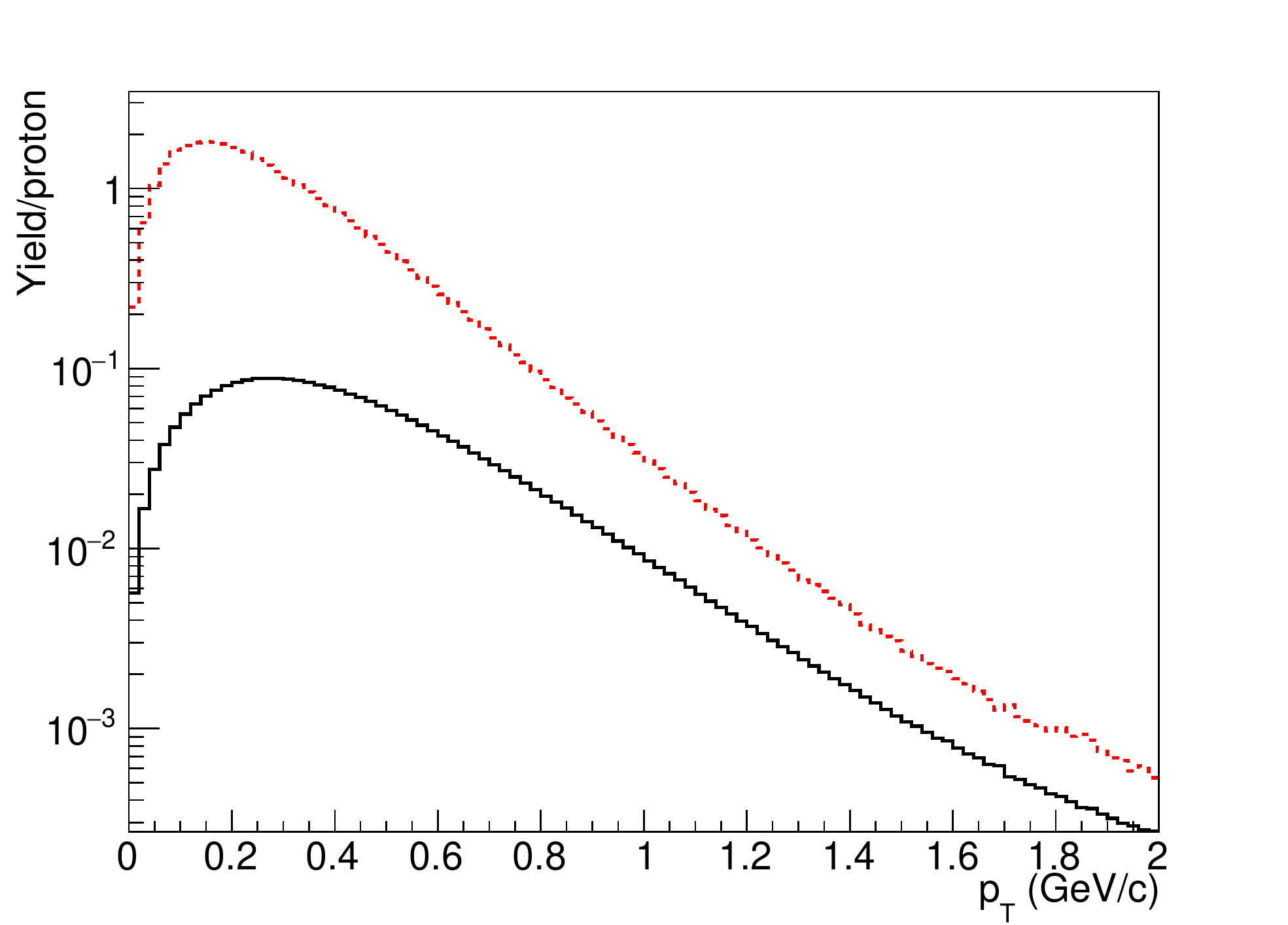}
\caption{\label{eta_momentum} The $\eta$ (solid) and $\pi^{0}$ (dotted) yield/proton as a function of the transverse momentum $p_{T}$ obtained from GEANT4 Monte Carlo simulations of 120 GeV protons interacting with the Fe beam dump.}\end{figure}


\begin{figure}
\includegraphics[clip,width=0.50\textwidth,angle=0]{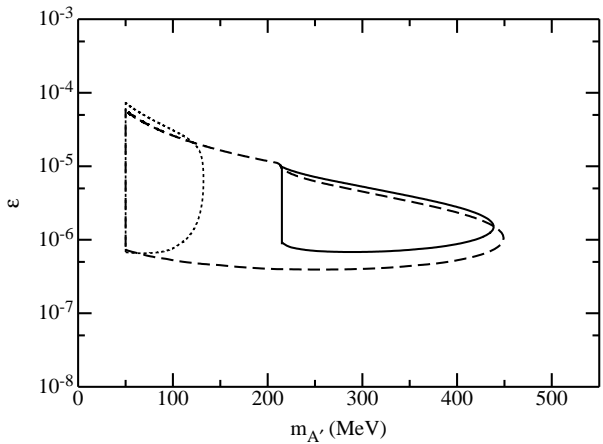}
\caption{\label{electron} Preliminary projections for an $\eta$ meson decaying to a photon ($\gamma$) 
and a dark photon, denoted by $A'$.  The $A'$
can subsequently decay into a $\mu^+\mu^-$  pair (solid), or an $e^+e^-$ pair (dashed). 
Here decays to the hidden sector do not occur.
The dotted curve represents the projected exclusion limit for $\pi^0$ decay. 
}
\end{figure}

\begin{figure}
\includegraphics[clip,width=0.50\textwidth,angle=0]{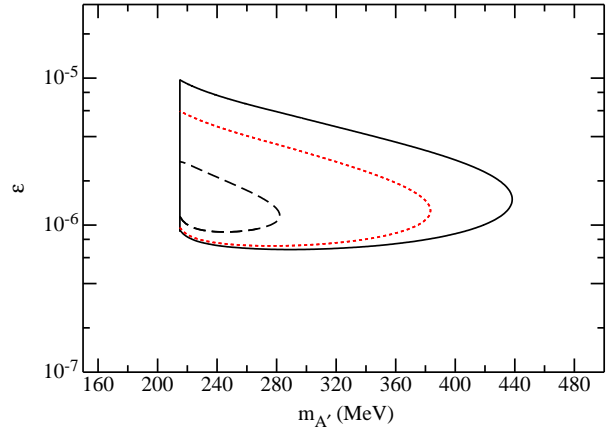}
\caption{\label{elecplushidden}
Here the $A'$ is produced by $\eta$ decay.  The $A'$ subsequently decays to a muon pair.
The preliminary projected exclusion regions for SeaQuest for three branching ratios to the dark sector between 0 and 90\% are shown.
Specifically, ${\cal B}(A'\to {\rm invisibles}=0,0.70,0.90$, which are shown as solid, dotted, and dashed 
curves, respectively.
}
\end{figure}

\begin{figure}
\includegraphics[clip,width=0.50\textwidth,angle=0]{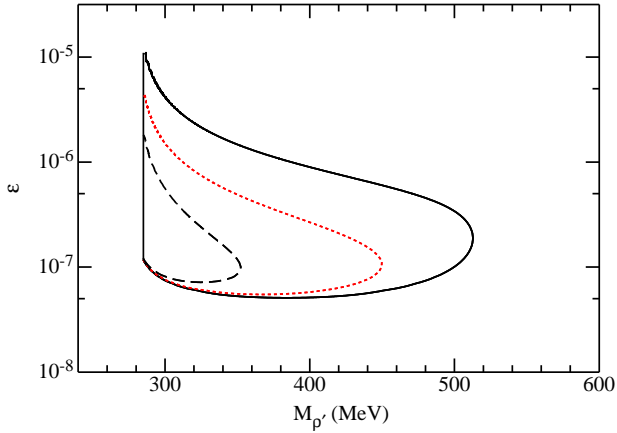}
\caption{\label{rho} Preliminary projections for an $\eta$ meson decaying to a photon
and a dark gauge boson $\rho'$ that couples to quarks,
with a probability governed by the square of the coupling constant between light and dark sectors of $\varepsilon^2$ with the probability
governed by Eq.~(\ref{decay2rhop}).
The $\rho'$
can subsequently decay into a charged pion pair with a decay lifetime proportional to $\varepsilon^{-2}$
governed by Eq.~(\ref{rhop2pi}). The $\rho'$ is permitted to decay to the dark sector with a branching ratio of
${\cal B}(\rho' \to {\rm invisibles}) = 0,0.80,0.95$, which are shown as solid, dotted, and dashed curves, respectively.
}
\end{figure}

In this projection, we assume that the fiducial region for detecting the decay of the dark photons to a muon pair begins 4-m downstream of the front edge of the beam stop and extends 0.85 m downstream.  
This fiducial region ensures that all decay muons must travel through at least 0.15 m of FMAG, which would impart an effective $p_T$ kick of about 0.08 GeV.  This ensures that the muon pair would be well separated before reaching the first detector station. For the decay of a $e^+e^-$ pair or in later discussion, a pion pair, the fiducial region is only 0.95 m and entirely outside the Fe magnet. 
For the case of $e^+e^-$ or $\pi^+\pi^-$ pairs, it might be necessary to add a small air gap magnet between the downstream end of FMAG and the first detector station to ensure a small $p_T$ kick.  For the purpose of simulation, it was assumed that the necessary separation of the pairs was met. To be specific, the expression used for the $A'$
decays to lepton pairs in the fiducial length, $l_{\rm fid}$, after the $A'$
traverses the length of the beam dump, $l_{\rm dump}$, is given by
\begin{eqnarray}
N_{\rm dec} &=& N_0 {\cal B}(A' \to \ell^+ \ell^-)
\exp\left(-\frac{l_{\rm dump}}{c\tau_{A'}}\frac{m_{A'}}{|{\mathbf p}_{A'}|}\right) \nonumber \\
&& \times \left[ 1 - \exp \left(-\frac{l_{\rm fid}}{c\tau_{A'}}\frac{m_{A'}}{|{\mathbf p}_{A'}|}\right)\right] \,,
\label{detect}
\end{eqnarray}
where ${\mathbf p}_{A'}$ is the $A'$ momentum in the laboratory frame, and $\tau_{A'}$ is the $A'$ lifetime in the $A'$ rest frame.

Two different approaches were used to determine the acceptance of the spectrometer for a dark photon. They both assumed a dark photon of various masses decaying at different vertex positions in the beam stop. The first method makes use of the full spectrometer geometry in a GEANT4 (version 4.9.6.p03~\cite{geant4ref}) 
simulation. The second method uses the spectrometer geometry in a simplified situation where multiple scattering and energy loss of the decay particles are put in explicitly. The kinematic variables and the dimuon yields were compared, and there is good agreement between the two approaches. Currently, our acceptance is limited by the decay length of the dark photon. Once track reconstruction is optimized and a detailed simulation of the background events are achieved, it may be possible to probe larger values in $\varepsilon$ parameter space.

The projected regions of sensitivity based on this simulation are indicated in Fig.~\ref{electron}.
The contours shown in the figure indicate 10-event contours, which should be adequate to set 95\% confidence levels on exclusion of dark photons.  For example, if we assume a signal of 10 events and a detected number of events of 10 with a background of 7 events, the confidence level \cite{Zech:1988un} that the true number of events is less than 10 is 94.6$\%$. Three cases are displayed:
(1) detection of a muon pair from $\eta$ decay to a dark photon and subsequent $A'$ decay to a muon pair,  
(2) detection of an $e^+e^-$ pair from $\eta$ decay and subsequent $A'$
decay, and (3) detection of $e^+e^-$ decay from neutral pion decay and subsequent $A'$ decay.  It is assumed that there are no decays to the invisible sector for the projections of Fig.~\ref{electron}.

Many projections and exclusion plots assume that the decay to the visible sector is 100\%.
In Fig.~\ref{elecplushidden} we explore the case where $A'$ decay to the invisible sector is also allowed
to determine its impact on the ability to limit the $A'$.
The figure shows the specific case of $\eta$ decay to $\gamma A'$, followed by $A' \longrightarrow \mu^+\mu^-$.
Furthermore, it is assumed that the branching ratio to the hidden sector, ${\cal B}(A'\to {\rm invisibles}$, varies from 0 to 90\%.
Clearly, if decay to the hidden sector is permitted, the region excluded by the experiments dramatically shrinks.

\subsection{$\eta$ meson decay to a charged pion pair}

With relatively small modifications, it may be possible for the pair spectrometer associated with the SeaQuest experiment to detect a charged pion pair.
Although the detectors in this spectrometer presently do
not have the capability to distinguish between an $e^+e^-$ pair and a pion pair, it may not be necessary in a first stage of the experiment.  If a mass peak is observed from a displaced vertex in the SeaQuest experiment, then the necessary particle identification could be added to the detector to determine whether the pair is leptonic or hadronic.  If we consider $\eta$ decay to a photon and a dark $\rho'$, and the dark $\rho'$ subsequently decays to a pion pair, the expected sensitivity plot is given in Fig.~\ref{rho}, again showing 10-event contours based on Eqs.~(\ref{decay2rhop}) and (\ref{rhop2pi}).
The effect of the $\rho'$ decay to the invisible sector has also been explored in the figure.

\section{Summary and Outlook}
\label{sec:sum}

In this paper we have explored the discovery prospects of the SeaQuest experiment, E906 at Fermilab, operated in
proton beam dump mode. We have explored the ability to which it can explore new regions of dark photon parameter
space, and how, if pion pair detection were made possible, it can search for an entirely
new kind of hidden sector,
one probed through the hidden strong interactions of quarks.
The latter scenario has never been explored in
beam dump experiments, and the possibility of strongly
coupled hidden sectors at sub-GeV mass scales
is poorly constrained by other considerations~\cite{Dobrescu:2014fca}.
We emphasize that the SeaQuest projectionss in this paper have been computed assuming one year of running with a
non-optimized detector. Increasing the exposure could lower the epsilon range considerably, and, in
the case of the proton bremsstrahlung mechanism, permit sensitivity to higher-mass gauge bosons as well.

Moreover, the prospect of polarized proton beams~\cite{Isenhower:2012vh}
opens a unique window of sensitivity to
light $Z'$ gauge bosons, to the $Z_d$
scenario of Refs.~\cite{Davoudiasl:2012ag,Davoudiasl:2012ig,Davoudiasl:2012qa,Davoudiasl:2014kua}
and, in the event of $\pi^+\pi^-$ detection,
to the leptophobic, light $Z'$ proposed in Ref.~\cite{Dobrescu:2014fca}.
Increasing either the proton beam energy or the length
of the fiducial region in the iron beam stop can also extend the sensitivity of SeaQuest to
heretofore unexplored regions of hidden-sector parameter space.

\begin{acknowledgments}
We greatly appreciate the assistance of M. M. Medeiros and B. Nadim with GEANT4 Monte Carlo simulations. We are grateful for useful discussions with C. Brown, D. Geesaman, P. Reimer, R. Gilman, J.-C. Peng,
R. Essig, J. Bl\"umlein, J. Brunner, and J. Qian. 
We are also grateful to P. Masjuan for helpful correspondence.
This work was supported by the Department of Energy (DOE),
Office of Science, Office of Nuclear Physics,
contract nos.~DE-FG02-96ER40989 (SG) and DE-AC02-06CH11357 (RH) and by the
National Science Foundation (NSF), grant no. NSF PHY 1306126 (AT).
\end{acknowledgments}

\bibliography{dark}

\end{document}